\documentclass[reprint,bibnotes,amsmath,amssymb,aps,prl,showpacs,floatfix,superscriptaddress]{revtex4-1}

\usepackage{amsmath}
\usepackage{amsfonts}
\usepackage{amssymb}
\usepackage{amsxtra}
\usepackage{xcolor}
\usepackage{graphicx}
\usepackage{subfigure}
\usepackage{dcolumn}
\usepackage{float}
\usepackage{bm}
\usepackage[breaklinks=true,colorlinks,citecolor=blue,linkcolor=blue,urlcolor=blue]{hyperref}

\newcommand{\nn}{\nonumber}

\newcommand{\bsigma}{\boldsymbol{\sigma}}
\DeclareMathAlphabet{\bi}{OML}{cmm}{b}{it}
\def\be{\begin{equation}}
\def\ee{\end{equation}}
\def\bearr{\begin{eqnarray}}
\def\eearr{\end{eqnarray}}
\def\la{\langle}
\def\ra{\rangle}

\begin{document}
\title{Nonlinear, anisotropic and giant photoconductivity in intrinsic and doped graphene}
\author{Ashutosh Singh}
\affiliation{Department of Physics, Indian Institute of Technology Kanpur, Kanpur - 208016, India}
\author{Saikat Ghosh}
\affiliation{Department of Physics, Indian Institute of Technology Kanpur, Kanpur - 208016, India}
\author{Amit  Agarwal}
\email{amitag@iitk.ac.in}
\affiliation{Department of Physics, Indian Institute of Technology Kanpur, Kanpur - 208016, India}

\date{\today}

\begin{abstract}
We present a framework to calculate the anisotropic and non-linear photoconductivity for two band systems with application to graphene. In contrast to the usual perturbative (second order in the optical field strength) techniques, we calculate photoconductivity to all orders in the optical field strength. In particular, for graphene, we find the photoresponse to be giant (at large optical field strengths) and anisotropic. The anisotropic photoresponse in graphene is correlated with polarization of the incident field, with the response being similar to that of a half-wave plate. We predict that the anisotropy in the simultaneous measurement of longitudinal ($\sigma_{xx}$) and transverse $(\sigma_{yx})$ photoconductivity, with four probes, offers a unique experimental signature of the photo-voltaic response, distinguishing it from the thermal-Seebeck and bolometric effects in photoresponse. 
\end{abstract}

\maketitle
A range of studies on photoconductivity (or, scanning photoresponse microscopy) in graphene \cite{Gabor648,LeeEduardo,Lemme,Park,Zhang,Freitag,Xu,Xia,Xia2,Tielrooij, Kim,Liu,Zhenhua,Sun1,PhysRevB.79.245430, Frenzel,Polini} have led to understandings, for example, on the nature of photo-excited carriers, hot carrier relaxation,  the impact of metal contacts and doping asymmetry. The bulk of these studies, either theoretically or experimentally, have explored the linear response regime, where the system response to the incident optical field can be treated perturbatively \cite{Schliemann}.  
However, it has also been shown that the non-linear response of the photo-excited carriers leads to several new phenomenon, modifying absorbtion, optical conductivity and even plasmons~\cite{Vasko,Mishchenko,Singh,Chaves}.  In this letter, we extend such non-linear studies to understand modifications in photoresponse, in particular, photovoltaic conductivity of two-band systems in general, and graphene in particular, due to non-linear and anisotropic steady state photo-excited carriers. 

\begin{figure}[t!]
\centering
\includegraphics[width=\linewidth]{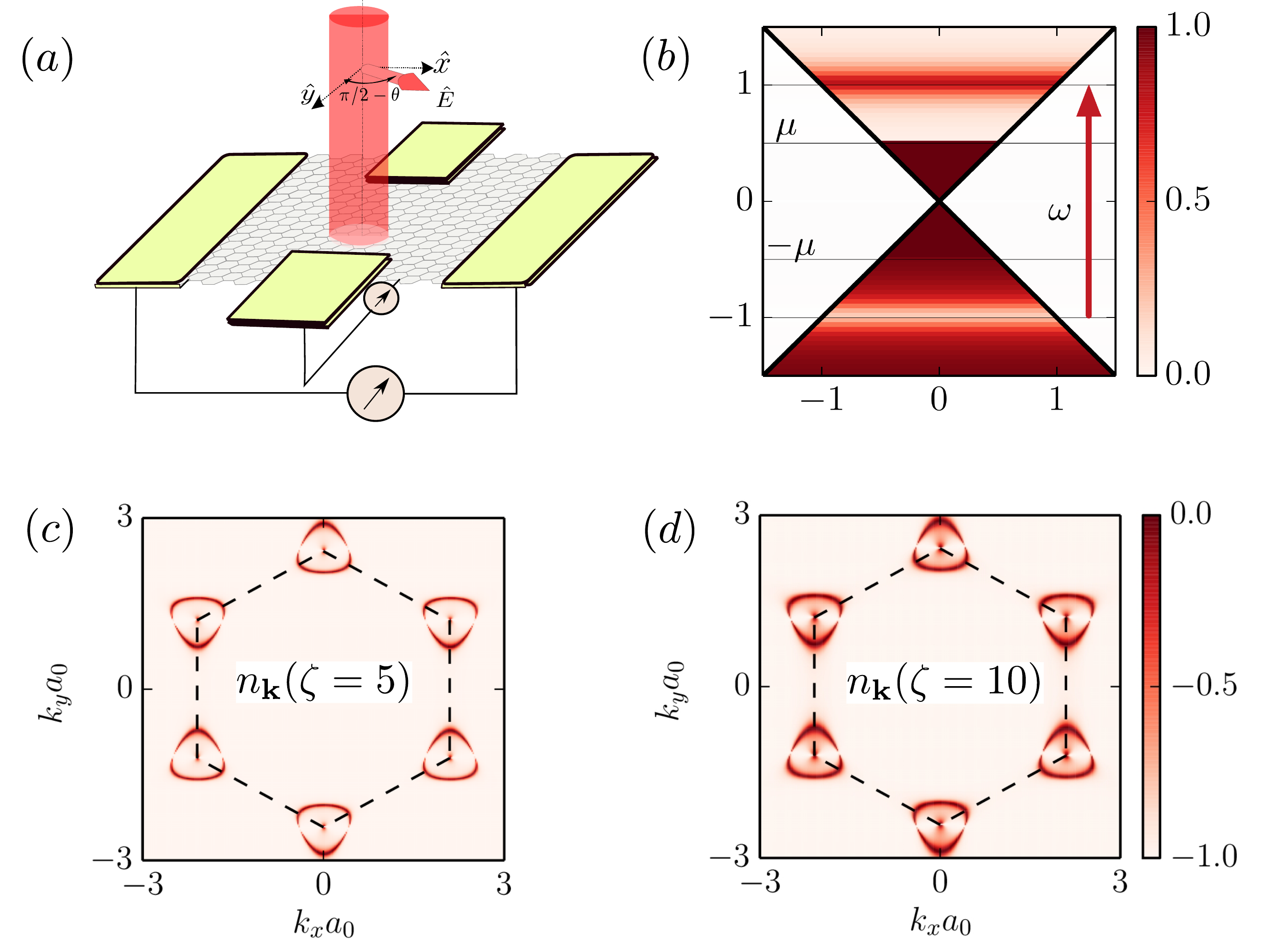}
\caption{ (a) Schematic of the proposed four-probe setup to measure longitudinal ($\sigma_{xx}$) and transverse ($\sigma_{yx}$) photoconductivity simultaneously. 
(b) Color plot of the photo-excited carrier NDF in the energy-$k_y$ plane, based on Eq.~\eqref{fc}, for $\zeta = 10$ and $\mu = \hbar \omega/2$. 
(c) and (d) show the momentum resolved population inversion in the full Brillouin zone, for $\zeta = 5$ and $10$, respectively for a $x$-polarized light in pristine graphene. The anisotropy of the carrier NDF function, and the increase in the  number of photo-excited carriers with increasing $\zeta$ is evident.}
\label{Fig1}
\end{figure}

Accordingly, we first develop a general formulation for estimating steady-state nonlinear photovoltaic conductivity in two band systems, when illuminated with a continuous wave (CW) optical field. The formulation goes beyond usual perturbative 
(quadratic in the optical field strength - $E_0$) calculations, obtaining an almost exact anisotropic non-equilibrium distribution function (NDF), in terms of the interband population inversion $n_{\bf k}$, to all orders in $E_0$.  While the NDF reduces to Fermi function in the $E_0 \to 0$ limit, it leads to a saturation regime for large fields, in the  $E_0\to \infty$ limit. Conveniently, a single parameter $\zeta ~\propto E_0 $ [see  Eq.~\eqref{eq:zeta}] characterizes the nonlinearity in the photo-excited NDF \cite{Mishchenko,Singh}. The impact of $\zeta$ on the NDF, along with the anisotropy of $n_{\bf k}$ is highlighted in panels (c) and (d) of Fig.~\ref{Fig1}. 
The steady state NDF, parametrized by $\zeta$, is then used in semi-classical Boltzmann transport formalism to calculate the photovoltaic conductivity. Within this formulation, we predict highly anisotropic (locked with the polarization of the optical field) and giant (at large fields) photovoltaic conductivity. 

In almost all scanning photo-response experiments, signatures of photovoltaic response overlap with other photo-induced physics, including thermal Seebeck effect (TSE), thermal bolometric (TBE) and photo-gating effects, contributing to photoconductivity many a times with identical signatures \cite{Lemme,Freitag,Polini}. 
To separate out the photo-voltaic contribution experimentally, we propose a unique distinctive signature: in steady state, peaks in longitudinal ($\sigma_{xx}$) and transverse ($\sigma_{yx}$) conductivity as a function of optical polarization angle, should be shifted by $\pi/4$ for the photo-voltaic contribution. The effect of thermal Seebeck effect can be subtracted out by a measurement of the photocurrent in absence of the dc bias, and the thermal bolometric effect is likely to be isotropic. Accordingly, simultaneous measurement of photo-conductivity with varying light polarization, in a four-probe scenario can reveal a pure photo-voltaic signature [see Fig.~\ref{Fig1}(a)]. Further scaling of the peaks with optical field strength, as we estimate here in the non-linear regime, can then better characterize and differentiate such competing signatures.

Our starting point is a generic two band system \cite{Singh}, whose quasiparticle dispersion is described by the Hamiltonian,
$
\hat H_0 = \sum_{\bf k}{\bf h}_{\bf k} \cdot \bsigma$,
where $ {\bf h}_{\bf k} = (h_{0{\bf k}},h_{1{\bf k}}, h_{2{\bf k}}, h_{3{\bf k}})$ is a vector composed of real scalar elements and $ \bsigma = (\openone_2, \sigma_x, \sigma_y, \sigma_z)$ is 
 a vector composed of the identity and the three Pauli matrices. Within the dipole approximation the dynamics of an electron in presence of an external electromagnetic field, is governed by the Hamiltonian\cite{Aversa}:     
$\hat H_{ \rm em} = \hat H_0 + e {\bf E}\cdot \hat{\bf r} $.  Here $\hat H_0$ is a diagonal matrix comprising of the dispersion of conduction ($\varepsilon^c_{\bf k}$) and valence ($\varepsilon^v_{\bf k}$) bands, 
and the optical matrix elements coupling the optical field to the carriers are defined by  
$e{\bf r}^{\lambda \lambda'} ({\bf k}) = e \la \psi^\lambda |\hat{\bf r}|  \psi^{\lambda'} \ra$, where $\lambda = {c (v)}$ denotes the conduction (valence) band. In this letter, we will focus only on vertical inter-band transitions, and ignore the finite momentum intra-band terms.
 
The momentum resolved optical Bloch equations for the inter-band population inversion $n_{\bf k} = \rho^{cc} - \rho^{vv} $, and 
the inter-band coherence $p_{\bf k} =  \rho^{vc} $ are given by \cite{Singh}
\bearr \label{COBE3a}
\partial_t {n}_{\bf k} & =& 4 \Im \left[ {\Omega}^{vc*}_{\bf k} {p}_{\bf k} \right] - \gamma_{1}({n}_{\bf k}-n^{\rm eq}_{\bf k})~, \\ 
\label{COBE3b}
\partial_{t}{p}_{\bf k} & = & i \omega_{\bf k}{p}_{\bf k}  - i {\Omega}^{vc}_{\bf k} {n}_{\bf k}-\gamma_{2}{p}_{\bf k}~,
\eearr
where $ \hbar \omega_{\bf k} = \varepsilon_{\bf k}^c - \varepsilon_{\bf k}^v$. In Eq.~\eqref{COBE3a}, $\hbar {\Omega}^{vc} = e{\bf E}\cdot{\bf r}^{vc}$, is the interband Rabi frequency
and ${\bf E}(t) = E_0 \cos(\omega t)\hat{\bf e}$, for a linearly polarized light with $\hat{\bf e} = (\cos\theta, ~\sin\theta)$  denoting the unit vector along the linear polarization direction of the 
incident optical field.  Further in Eq.~\eqref{COBE3a} $n^{\rm eq}_{\bf k} = f_{c\bf k}^{(0)} - f_{v\bf k}^{(0)}$  where $f_{\lambda{\bf k}}^{(0)}= [1+ \exp{[(\epsilon_{\bf k}^\lambda - \mu)/k_B T]}]^{-1}$ 
denotes the Fermi function with band index $\lambda$ and $\gamma_{1}$, $\gamma_{2}$ are the phenomenological 
relaxation rate for the population inversion and  the inter-band coherence, respectively. In general $\gamma_1$ 
and $\gamma_2$ are $\omega$ and ${\bf k}$ dependent, however for simplicity we will assume them to be constants. Moreover, 
their energy/momentum dependence can be easily included in the mentioned formalism, and it does not change the results in any qualitative way.

For a continuously incident optical wavefront, in the steady state regime,  Eqs.~\eqref{COBE3a}-\eqref{COBE3b} can be solved to
obtain the following momentum resolved steady state value of the conduction band and valence band NDF:  
\bearr\nn\label{fc}
\rho^{cc}  \equiv f_{c \bf k}^{(1)} &=& \frac{1}{2}\left[f_{c \bf k}^{(0)}(1 + G_{\bf k}) + f_{v \bf k}^{(0)}(1 - G_{\bf k})\right],\\
\rho^{vv}  \equiv f_{v \bf k}^{(1)} &=& \frac{1}{2}\left[f_{c \bf k}^{(0)}(1 - G_{\bf k}) + f_{v \bf k}^{(0)}(1 + G_{\bf k})\right].
\eearr
%
Here 
\be \label{G}
 G_{\bf k} = \left[ 1 + \zeta^2~\frac{\omega^2}{\omega_{\bf k}^2} ~ \frac{2\gamma_2^2|\tilde {\bf M}^{vc}\cdot
 \hat{\bf e}|^2(\omega^2 + \omega_{\bf k}^2 + \gamma_2^2)}
 {\left[(\omega_{\bf k}^2 - \omega^2)^2 + 2\gamma_2^2(\omega_{\bf k}^2 + \omega^2) + \gamma_2^4\right]}\right]^{-1},
\ee
is a momentum dependent function, dictating the steady state population inversion, while 
$\tilde {\bf M}^{vc} = {\bf M}^{vc}/(ev_F)$ is a dimensionless, material dependent, optical-dipole matrix element, written in momentum gauge\cite{Singh,Aversa}. See supplementary material \footnote{\href{https://www.dropbox.com/s/myvbc89qgpz925k/Supple_v1.pdf?dl=0}{See supplementary material}} for the details of the calculation. The  dimensionless parameter $\zeta$ in Eq.~\eqref{G}, of the form, \cite{Mishchenko,Singh}
\be \label{eq:zeta}
\zeta \equiv \frac{e E_0 v_F}{\hbar \omega \sqrt{\gamma_{1} \gamma_{2}}}~,
\ee
quantifies the degree of non-linearity in the system.
In the limiting case of vanishing optical field intensity  ($\zeta \to 0$ limit), we have  
$G_{\bf k} \to 1$ and it leads to $f_{\lambda \bf k}^{(1)} \to f_{\lambda \bf k}^{(0)}$, restoring the normal ground state of the electron gas with $\rho^{cc} \to f_{c \bf k}^{(0)}$ and $\rho^{vv} 
\to f_{v \bf k}^{(0)}$. The opposite limit of large optical field intensity ($\zeta \to \infty$), characterizes a saturation regime, with equal number of carriers in 
the valence and conduction band, such that $n_{\bf k} \equiv (f_{c \bf k}^{(1)}- f_{v \bf k}^{(1)})\to 0$~.  

The photo-excited population inversion has two important aspects: 1) increasing $E_0$ increases $n_{\bf k}$ for optical energies $\hbar \omega > 2 \mu$ [see Fig.~\ref{Fig1}(b)-(d)], leading to  giant photoconductivity, and 2) the optical dipole matrix is inherently anisotropic and that is also reflected in the photo-excited NDF [see Fig.~\ref{Fig1}(c)-(d)]. This anisotropic NDF, on being   
`driven' by the external dc bias field, leads to anisotropic photovoltaic conductivity.

Using the steady state photo-excited NDF, we now calculate the response of the carriers to an applied bias field or equivalently,  photovoltaic conductivity.  
In a semi-classical scenario, for a $d$ dimensional system, to lowest order in the bias field, the photoconductivity 
(for a given $\omega$ and $E_0$) is given by  \cite{ashcroft1976solid} 
\be \label{eq:drude}
\sigma_{ij}= -\frac{g_s e^2}{{\gamma_{0}-i\omega_0}}\int_{\rm BZ}\frac{d\bf k}{(2\pi)^d} \sum_{\lambda = v,c} \frac{1}{\hbar^2}
\frac{\partial\varepsilon_{\lambda \bf k}}{\partial k_i}
\frac{\partial\varepsilon_{\lambda \bf k}}{\partial k_j}\frac{\partial f_{\lambda \bf k}^{(1)}}{\partial\varepsilon_{\lambda \bf k}}~,
\ee
where  $f_{\lambda \bf k}^{(1)}$ is now the non-equilibrium photo-excited NDF specified by Eq.~\eqref{fc},
with $\omega_0(\ll\omega)$ is the frequency of the applied bias and $g_s$ denotes the spin degeneracy. In Eq.~\eqref{eq:drude} $\gamma_{0}$ is the elastic momentum relaxation rate, which, primarily being dominated by impurity scattering, is assumed to be a constant in a mean-fled sense. Furthermore, $\sigma_{ij} = \sigma_{ji}$.
It can be noted that equation~\eqref{eq:drude} is simply a {\it non-linear} generalization (in the optical field strength) of the Drude conductivity, at finite frequency, for a two band system.  For the zero field limit, we get, $\lim_{\zeta \to 0}\sigma_{ii} = \sigma_{\rm Drude}$.  
In particular, for a continuum model of graphene, we get back the well known form, $\sigma_{\rm Drude} \equiv \sigma_D = e^2 |\mu|/(\pi \hbar^2 \gamma_0)$, where $\mu$ denotes the chemical potential. 

For particle hole symmetric systems, such as graphene, Eq.~\eqref{eq:drude} simplifies to 
\be \label{eq:drude2}
\sigma_{ij}(\omega)= -\frac{g_s (e^2/\hbar)}{\gamma_{0}-i\omega_0} \int_{\rm BZ}\frac{d\bf k}{(2\pi)^d} \frac{1}{2}
\frac{\partial\omega_{\bf k}}{\partial k_i}\frac{\partial\omega_{\bf k}}{\partial k_j} \frac{\partial 
{n}_{\bf k}}{\partial\omega_{\bf k}}~.
\ee
It is useful to express Eq.~\eqref{eq:drude2} as a sum of two parts with differing physics:  
$\partial_{\omega_{\bf k}} {n_{\bf k}}  = G_{\bf k} \partial_{\omega_{\bf k}}n^{\rm eq}_{\bf k} +  n^{\rm eq}_{\bf k} F_{\bf k}$, 
where $F_{\bf k} \equiv \partial_{\omega_{\bf k}}G_{\bf k}$. In this, the first term involves the derivative of the Fermi function, accounting for changes in carriers in vicinity of the Fermi surface. On the contrary, the second term is due to the  photo-excited carriers beyond the Fermi energy, if $\hbar \omega > 2 \mu$. The conductivity components corresponding to these two terms will be 
denoted as $\sigma^{(1)}_{ij}$ and $\sigma^{(2)}_{ij}$, respectively, hereafter.   
In the limiting case of vanishing optical field,  $\zeta \to 0$, we have $G_{\bf k}\to 1$ and $F_{\bf k} \to 0$ and hence $\sigma^{(2)}_{ij} 
\to 0$ as expected. In the other limit of $\zeta \to \infty$ (saturation limit), $G_{\bf k}\to 0$ and becomes increasingly broad leading to $F_{\bf k} \to 0$, and consequently $\sigma^{(2)}_{ij} 
\to 0$. We emphasize here that typical perturbative photoconductivity calculations include only the $E^2$ term of $\sigma^{(2)}$, in the infinite coherence time or $\gamma_2 \to 0$ limit. See supplementary material \cite{Note1},  
for a reproduction of the perturbative calculations of Ref.~\cite{Schliemann} from our formalism. 

\begin{figure}[t!]
\includegraphics[width=\linewidth]{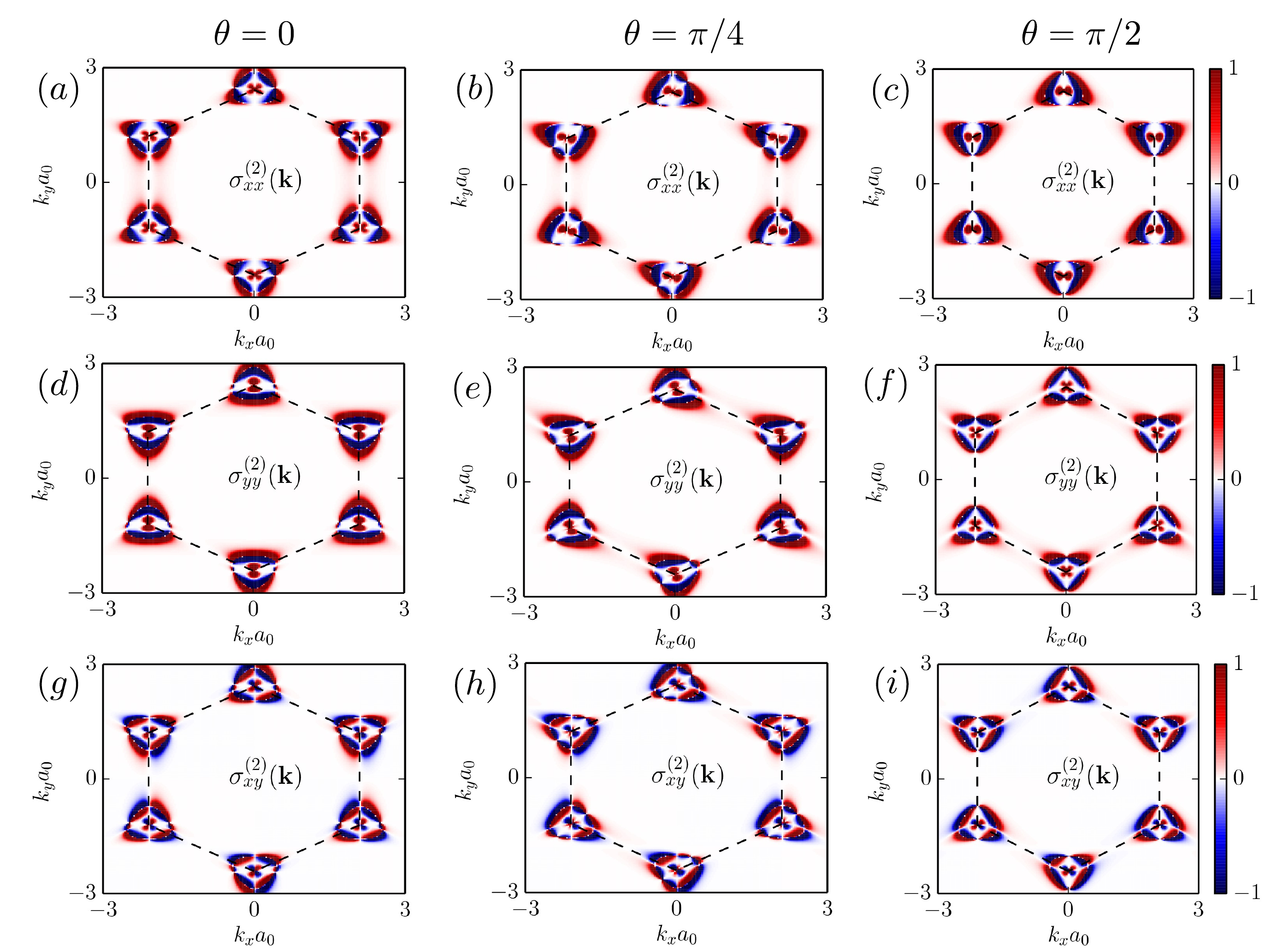}
\caption{(a), (b), (c) show the momentum resolved integrand of $\sigma^{(2)}_{xx}$ for the incident field polarization angles of $\theta = 0$, $\pi/4$  and $\pi/2$, respectively. Similarly (d), (e), (f) 
show the momentum resolved integrand of $\sigma^{(2)}_{yy}$ and, (g), (h), (i) show the momentum resolved integrand of
$\sigma^{(2)}_{xy}$
for the same set of $\theta$'s. In all panels we have fixed $\zeta=10$, and $\mu = 0$. 
Other parameters are: $\gamma_1 = 10^{12}s^{-1}$, $\gamma_2 = 5\times10^{13}s^{-1}$ and $\omega = 5\times10^{15}s^{-1}$. 
The dashed black lines show the Brillouin zone boundary of graphene.  This `rotation' of the integrand with change of the polarization direction of the incident optical field is what leads to anisotropic photoconductivity.  
}
\label{Fig2}
\end{figure}

The tight-binding Hamiltonian of graphene is given by 
$H = t\sum_{\langle i,j\rangle,s}(a^{\dagger}_{s,i}b_{s,j} + \text{H.c.})$, with $a_{s,m}$ and $b_{s,m}$ being the
annihilation operator for an electron in different sublattices at the lattice point $m$, and $s$ refers to the spin index. The hopping parameter $t \approx 2.7$ eV. Transforming to momentum space and 
casting it in the form of $\hat H_0 = \sum_{\bf k}{\bf h}_{\bf k}\cdot \bsigma$,  we have $h_{0 \bf k} = h_{3 \bf k} = 0$ and 
\bearr
h_{1\bf k} &=& t\left(\cos{k_x} + 2\cos{\frac{k_x}{2}}\cos{\frac{\sqrt{3}k_y}{2}}\right),\\
h_{2\bf k} &=& t\left(\sin{k_x} - 2\sin{\frac{k_x}{2}}\cos{\frac{\sqrt{3}k_y}{2}}\right).
\eearr
Here $k_{x/y}$ denote the dimensionless Bloch wave-vectors. 

The real component of the giant anisotropic photovoltaic conductivity (at zero temperature for $\omega_0 = 0$) is given by $\sigma_{ij} = \sigma_{ij}^{(1)} + \sigma_{ij}^{(2)}$, where
\bearr\nn\label{sigma_gr}
\frac{\sigma_{ij}^{(1)}}{\alpha\sigma_0} &=& \int_{\rm BZ}d{\bf k}~G_{\bf k}~\frac{\partial\tilde\varepsilon_{\bf k}}{\partial{k_i}}
\frac{\partial\tilde\varepsilon_{\bf k}}{\partial{k}_j}~\delta(|\tilde\mu|-\tilde\omega_{\bf k}/2).\\
\frac{\sigma_{ij}^{(2)}}{\alpha\sigma_0} &=& 2\int_{\rm BZ}d{\bf k}~\frac{\partial G_{\bf k}}{\partial \tilde\omega_{\bf k}}
~\frac{\partial\tilde\varepsilon_{\bf k}}{\partial{k_i}} \frac{\partial\tilde\varepsilon_{\bf k}}{\partial{k}_j}~\Theta(\tilde\omega_{\bf k}/2-|\tilde\mu|). 
\eearr 
where $\alpha = 2t^2/(\pi^2\hbar^2\omega\gamma_0)$, $\sigma_0 = e^2/(4\hbar)$ is the universal optical conductivity of graphene, $\tilde\varepsilon_{\bf k} = \varepsilon_{\bf k}/t$,
$\tilde\mu = \mu/(\hbar\omega)$ and $\tilde\omega_{\bf k} = \omega_{\bf k}/\omega$.
The explicit form of the optical matrix elements and the related velocities appearing in Eq.~\eqref{sigma_gr}, are given in the Supplementary material \cite{Note1}. An interesting case is that of intrinsic graphene ($\mu =0$), for which $\sigma_{ij}^{(1)}$ vanishes and the full contribution to photovoltaic conductivity comes from the photo-excited carriers only. 

The momentum dependent integrand for $\sigma_{ij}^{(2)}$ in Eq.~\eqref{sigma_gr} is shown in Fig.~\ref{Fig2} for the longitudinal ($\sigma_{xx}^{(2)}$ and $\sigma_{yy}^{(2)}$ ) as well as the transverse ($\sigma_{xy}^{(2)}$)
photoconductivity for three different polarization angles $\theta$, of the linearly polarized optical field. The change in the integrand, with changing $\theta$ for $\sigma_{xx}^{(2)}$, $\sigma_{yy}^{(2)}$ and $\sigma_{xy}^{(2)}$ is evident. The origin of this anisotropy is the $\theta$ dependence of the non-linear population inversion function $G_{\bf k}$, via the term 
$|{\bf M}_{\bf k}^{vc}\cdot\hat{\bf e}| = |M_{x{\bf k}}^{vc}\cos\theta + M_{y{\bf k}}^{vc}\sin\theta|$ [see Eq.~\eqref{G}].
This anisotropy is better understood by using the low energy Hamiltonian of graphene: $H = \hbar v_F \bsigma \cdot {\bf k}$. From Eq.~[15] of \cite{Note1}, we have ${\bf M}_{\bf k}^{vc}\cdot\hat{\bf e}\propto i\sin(\phi_{\bf k}-\theta)$, with $\phi_{\bf k}$
being the azimuthal angle for vector ${\bf k}$. This immediately implies that the number of
photo-excited carriers is minimum (maximum) along (transverse to) the direction of polarization of the optical field, $\hat{\bf e}$.
Thus an external dc bias field applied along (transverse to) the polarization field has low (high) number of photo-excited carriers to contribute to the photocurrent. As a consequence it is expected that in graphene, orthogonal bias and polarization fields should 
lead to maximal longitudinal photoconductivity \cite{Schliemann}. For example, $\sigma_{yy}$ ($\sigma_{xx}$) should be maximum (minimum) 
for $ \hat{\bf e} = \hat{x}$ (or $\theta =0$). 

\begin{figure}[t!]
\begin{center}
\includegraphics[width = \linewidth]{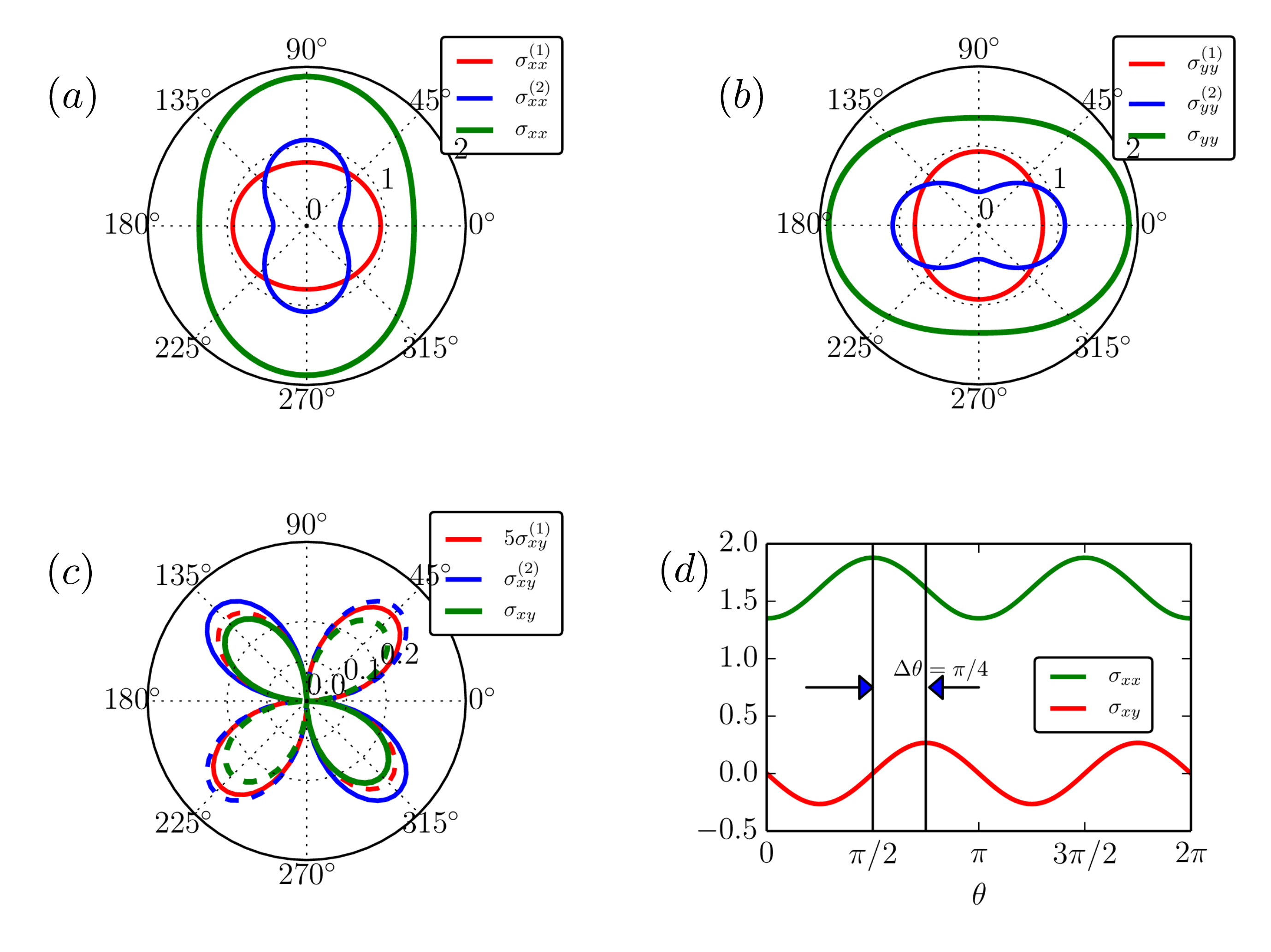}
\caption{ (a) $\sigma_{xx}$, (b) $\sigma_{yy}$ and (c) $\sigma_{xy}$ in units of $\sigma_{D}$, as a function of the electric field polarization angle $\theta$ for $\zeta = 1$, along with $\sigma_{ij}^{(1)}$ and $\sigma_{ij}^{(2)}$. $\sigma_{xx}$ and $\sigma_{yy}$ are simply $\pi/2$ rotated with respect to each other, as expected. $\sigma_{xy}$ changes sign as a function of the polarization angle, and this is indicated via the dashed (solid) line showing negative (positive) values in panel (c). The anisotropy in photoconductivity as a function of the direction of the linearly polarized light, can be used as a distinguishing qualitative feature of the photo-excited carriers from other thermal and bolometric effects: d) A simultaneous measurement of $\sigma_{xx}$ and $\sigma_{yx}$ as a function of $\theta$ will show peaks and troughs which differ by a factor of $\pi/4$. {For these plots we have used: $\omega = 10^{15}s^{-1}$, $\mu = 0.1 \hbar \omega$, $\gamma_1 = 10^{12}s^{-1}$ and $\gamma_2 = 5\times10^{13}s^{-1}$.}}  
\label{Fig3}
\end{center}
\end{figure}

This is indeed the case, as highlighted in Fig.~\ref{Fig3} which shows the polar plot of $\sigma_{xx}~, \sigma_{yy},~{\rm and}~\sigma_{xy}$ as a function of the polarization angle for a fixed value of the incident field strength. As expected $\sigma_{xx}$ is minimum ($\sigma_{yy}$ is maximum) for $\theta =0$ and maximum ($\sigma_{yy}$ is minimum) for $\theta = \pi/2$.
The angular dependence of the conductivity arising from the photo-excited carriers is given by \cite{Note1}, $\sigma_{xx}^{(2)} \propto (2 - \cos2 \theta)$, $\sigma_{yy}^{(2)} \propto (2 + \cos2 \theta)$, and $\sigma_{yx}^{(2)} \propto - \sin2 \theta$. While this suggests the ratio of the maximum to minimum longitudinal conductivity to be 3:1 \cite{Schliemann}, the angular dependence of $\sigma^{(1)}_{ii}$ is actually field dependent as shown in Eq.~[23] of \cite{Note1}. Note that the polarization angle dependence of the conductivity tensor is analogous to that of the rotation of the polarization vector by an angle $2 \theta$ by a half-wave plate in optics, on rotating the polarization angle by $\theta$.

This different anisotropy in $\sigma_{xx}$ and $\sigma_{yx}$ has remarkable experimental consequences. It offers a unique qualitative signature to distinguish the photo-excited carrier based photoconductivity from other bolometric and thermal effects. A {\it simultaneous measurement} of $\sigma_{xx}$ and $\sigma_{yx}$ (or $\sigma_{yy}$ and $\sigma_{xy}$), as a function of the polarization angle of the incident light, should have corresponding maxima's differing by phase of $\pi/4$ [see Fig.~\ref{Fig3}(d)]. An additional experimental check is that there will be no $\sigma_{xy}$ and no anisotropy in the directional conductivities for a circularly polarized light as shown in the supplementary material \cite{Note1}.  

\begin{figure}[t!]
\includegraphics[width=\linewidth]{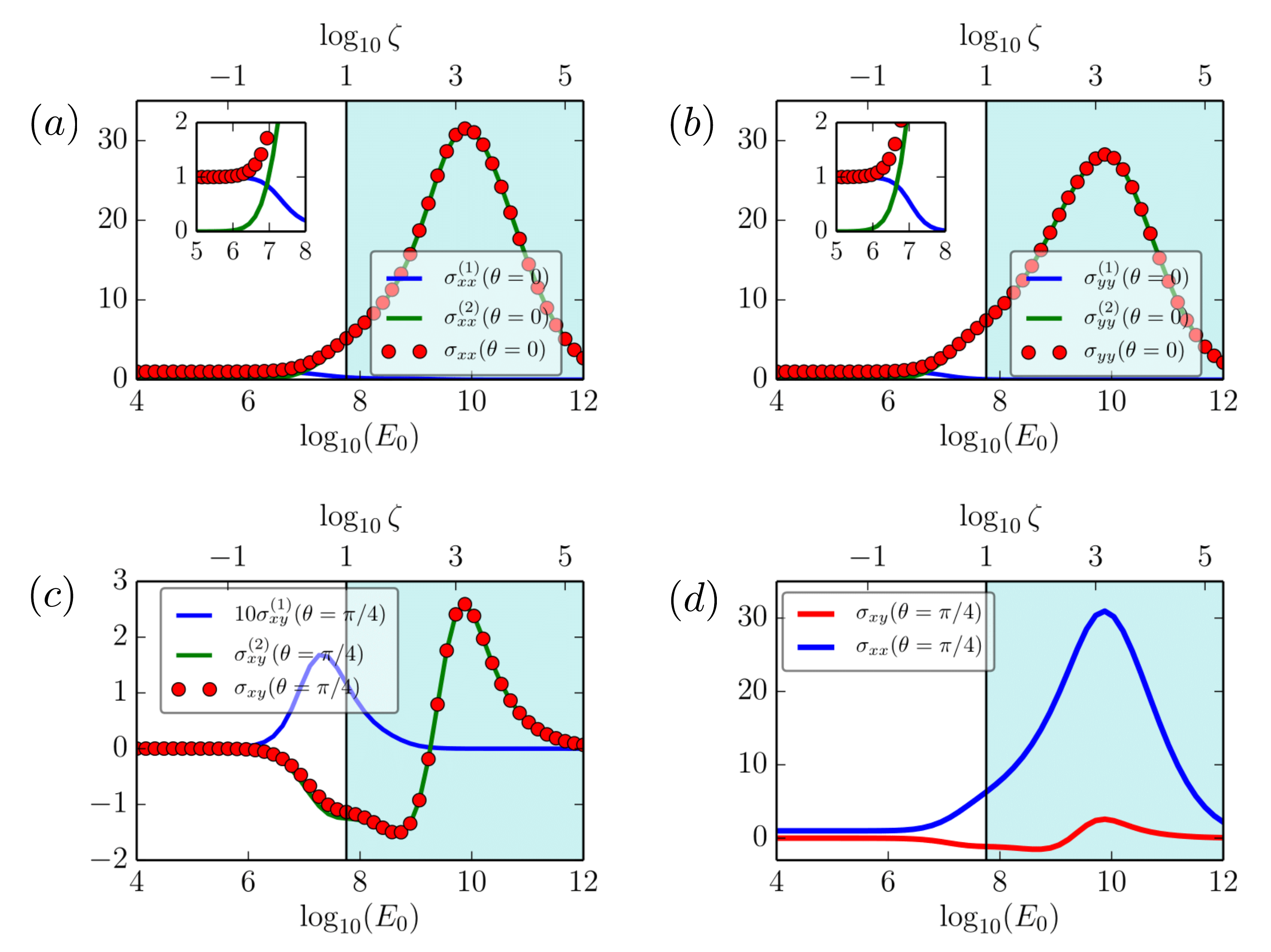}
\caption{(a) $ \sigma_{xx}$ and (b) $\sigma_{yy}$ in units of $\sigma_{D}$, as a function of incident optical field strength for $\theta = 0$. 
Starting with the Drude conductivity value for smaller values of the incident field strength, both $\sigma_{xx}$ and $\sigma_{yy}$ increase with increasing electric field till saturation of the population inversion is achieved, and then 
decrease on account of the increased broadening of the photo-excited NDF. 
In panel $(c)$ we have shown $\sigma_{xy}$ along with $\sigma_{xy}^{(1)}$ and $\sigma_{xy}^{(2)}$ 
as a function of optical field strength (in units of V/m) for $\theta = \pi/4$. 
The shaded region marked in all panels marks the $\zeta >1$ region. 
Other parameters used in this plot are the same as in Fig.~\ref{Fig3}.} 
\label{Fig4}
\end{figure}

The giant photovoltaic conductivity is a consequence of the large number of photo-excited carriers centered 
around energies $\varepsilon^c_{{\bf k}}=\hbar \omega/2$ (for $\hbar \omega > 2 \mu$). The number of such carriers 
increases with increasing field strength [recall that $G_{\bf k} \propto 1/(1+E^2 f_{\bf k})$], till saturation  of the inverted population sets in ($\rho^{cc}=\rho^{vv}$ for $\zeta \to \infty$ limit). {The giant photoconductivity is expected to be even larger in systems with parabolic dispersion due to larger velocity 
of the higher energy photo-excited carriers. }

Figure~\ref{Fig4} shows the dependency of $\sigma_{xx}$, $\sigma_{yy}$ and $\sigma_{xy} = \sigma_{yx}$ on the optical field 
strength. For small field strength, $\sigma_{xx} =\sigma_{yy} = \sigma_{\rm D}$, and increase with increasing electric field strength as $\zeta >1$, and the number of photo-excited carriers becomes significant.  
Note that $\sigma_{ii}^{(1)}$ gets contribution from the sharpness of the Fermi surface, 
and as the number of photo-excited carriers increase, it reduces the sharpness of the Fermi surface and thus 
$\sigma_{ii}^{(1)} \to 0$ for $\zeta \ge 4 \mu/(\hbar \gamma_2)$ when $\omega \gg \gamma_2$ and $\omega \gg \mu$ (see Eq.~(23) in \cite{Note1}). Unlike the longitudinal conductivity, the transverse conductivity $\sigma_{xy} = \sigma_{yx}$ increases from zero and its sign depends on the polarization angle. Here the dominant contribution primarily comes from the photo-excited carriers as $\sigma^{(1)}_{xy} \ll \sigma^{(2)}_{xy}$ -- see Fig.~\ref{Fig4}(c). As a consequence we have $\sigma_{xy} \propto - \sin(2 \theta)$. 

 Another interesting fact is that the broadening of $G_{\bf k}$ also increases with increasing $\zeta$ -- see Eq.~\eqref{G}. This leads to $G_{\bf k}$ becoming almost flat and very close to 0, and $F_{\bf k} \to 0$ for $\zeta \to \infty$.  
As a consequence of this increased broadening of the photo-excited NDF for large $\zeta$ we have $\sigma^{(2)}_{ij} \to 0$. This is reflected in the fact that all three photovoltaic conductivities $\sigma^{(2)}_{xx}$, $\sigma^{(2)}_{yy}$ and $\sigma^{(2)}_{xy}$ first increase with increasing $\zeta$ and then peak around the same $\zeta$ value and eventually decrease to $0$ for $\zeta \to \infty$ in the saturation regime. 

Photoconductivity measurements in graphene have contributions from primarily three different physical phenomena: 1) photovoltaic effect - as discussed in detail in this article, (2) thermoelectric Seebeck effect - which is significant near the charge neutrality point,  and (3) thermal bolometric effects which are important away from the charge neutrality point \cite{Freitag, Polini}. 
TSE arises from the response of the photo-excited NDF to the temperature gradient generated by the incident field across its transverse profile and is present even in the case of zero bias. Thus the TSE contribution (which can in principle be anisotropic) to the photocurrent  can therefore be subtracted out from the total photocurrent leading to 
$\sigma_{ij}(E_0) = j^{\rm eff}_i/E^{\rm bias}_j$, where  $j^{\rm eff}_i=j_i(E^{\rm bias}_j,E_0) - j_i(0,E_0)$. This removes the impact of TSE in photoconductivity. The TBE is based on increased local temperature under the beam spot, and is expected to be isotropic, independent of the  polarization angle $\theta$. The phase lag of $\pi/4$ between the consecutive minima (maxima) of the two conductivity measurement [See Fig.~\ref{Fig3}(d)] will therefore be a unique experimental signature of photoconductivity arising from purely the photo-excited carriers.

To summarize, we have presented an analytical framework to calculate the steady-state, nonlinear photoconductivity for two band systems. 
Our framework goes beyond the usual perturbative calculations, using an almost exact photo-excited NDF in a semiclassical transport formalism to obtain non-linear, anisotropic photoconductivity. We find two distinct photo-voltaic contributions to the conductivity, arising from a) the carriers in
vicinity of the Fermi-surface,  and b) the photo-excited carriers centered around energies $\hbar \omega/2$. For graphene, we find the resulting photoconductivity to be giant (at large optical field strengths), and anisotropic with the anisotropy locked to the angle of polarization of the  optical field. Accordingly, we propose a simultaneous measurement of the steady state $\sigma_{xx}$ and $\sigma_{xy}$ in a four probe setup in graphene, for a unique experimental signature of this calculated photo voltaic effect, thereby differentiating it from other photo-thermal effects which are  expected to be either isotropic 
or can be easily subtracted out. 

\bibliography{refs_PC_graphene.bib}

\end{document}